\documentclass[sigconf,nonacm]{acmart}
\usepackage{booktabs} 
\usepackage{wrapfig} 
\usepackage{balance}
\usepackage{enumitem}
\usepackage{moresize}
\usepackage{subcaption}
\usepackage{epstopdf}
\usepackage{graphicx}
\usepackage{geometry} 
\usepackage[textsize=scriptsize]{todonotes}
\usepackage{marginnote}
\usepackage{color}
\usepackage{multirow} 
\usepackage{enumitem}
\usepackage{bm}
\usepackage{xcolor}
\usepackage[ruled, lined, linesnumbered, commentsnumbered, longend]{algorithm2e}
\usepackage{algorithmicx}
\usepackage{algpseudocode}
\usepackage{array}
\usepackage{soul}
\usepackage{makecell}
\usepackage[]{caption}
\usepackage{hyperref}
\usepackage{wrapfig}
\usepackage{amsmath}
\usepackage{textcomp}

\newcommand\vldbdoi{XX.XX/XXX.XX}
\newcommand\vldbpages{XXX-XXX}
\newcommand\vldbvolume{XX}
\newcommand\vldbissue{XX}
\newcommand\vldbyear{XXXX}
\newcommand\vldbauthors{\authors}
\newcommand\vldbtitle{\shorttitle} 
\newcommand\vldbavailabilityurl{https://github.com/BU-DiSC/SilentZNS-bench}
\newcommand\vldbpagestyle{plain}

\SetCommentSty{mycommfont}

\definecolor{myGreen}{rgb}{0.5, 0.8, 0.6}
\definecolor{myRed}{rgb}{0.83, 0.51, 0.58}
\definecolor{myBlue}{rgb}{0, 0.44, 1}
\definecolor{myGrey}{rgb}{0.57, 0.64, 0.69}

\newcommand{\PreserveBackslash}[1]{\let\temp=\\#1\let\\=\temp}
\newcolumntype{C}[1]{>{\PreserveBackslash\centering}p{#1}}
\newcolumntype{R}[1]{>{\PreserveBackslash\raggedleft}p{#1}}
\newcolumntype{L}[1]{>{\PreserveBackslash\raggedright}p{#1}}

\newcommand\Paragraph[1]{\vspace{0.02in}  \noindent \textbf{#1.}}
\newcommand\Paragraphnp[1]{\vspace{0.02in}  \noindent \textbf{#1}}

\newcommand\SubParagraph[1]{\vspace{0.02in}  \noindent \textit{\underline{\smash{#1.}}}}
\newcommand\SubParagraphno[1]{\vspace{0.02in}  \noindent\textit{\underline{\smash{#1 }}}}

\newcommand{\approach}{SilentZNS}

\begin{document}

\title{Eliminating the Hidden Cost of Zone Management in ZNS SSDs}


\author{Teona Bagashvili}
\affiliation{\institution{Boston University}
\country{}
}
\email{teona@bu.edu}

\author{Tarikul Islam Papon}
\affiliation{\institution{UMass Boston}
\country{}
}
\email{t.papon@umb.edu}

\author{Subhadeep Sarkar}
\affiliation{\institution{Brandeis University}
\country{}
}
\email{subhadeep@brandeis.edu}

\author{Manos Athanassoulis}
\affiliation{\institution{Boston University}
\country{}
}
\email{mathan@bu.edu}



\begin{abstract}
Zoned Namespace (ZNS) SSDs offer a new storage model that allows for high throughput and low-latency storage by eliminating device-side garbage collection.
The ZNS interface exposes storage as \textit{append-only zones}, thus enforcing host applications (e.g., database systems) to append, read, and garbage collect their pages.
However, the storage abstraction of ZNS SSD hides the substantial differences across different ZNS SSD controller designs, which affects both the performance and predictability of host applications.
We find that existing ZNS controllers exhibit (a) increased \textit{device-level write amplification} (DLWA), (b) increased \textit{wear}, and (c) increased \textit{interference with host I/O}.
We identify that (i) zone allocation granularity, (ii) zone geometry, (iii) write order, and (iv) zone mapping and management strategy are the four main causes behind this.

To provide a predictable storage device, we propose \textit{\approach{}}, a new holistic zone management approach that expands the design space of zones and allocates blocks to zones on the fly, while minimizing wear, maintaining parallelism, and avoiding superfluous writes to the device.
\approach{} is a flexible zone allocation scheme that departs from traditional logical-to-physical zone mapping and allows arbitrary collections of blocks to be assigned to a zone. 
\approach{} further guarantees wear-leveling and competitive read performance, while substantially reducing DLWA.
We implement \approach{} using the state-of-the-art ConfZNS++ emulator and evaluate it on synthetic microbenchmarks and key-value storage engines.
We show that \approach{} reduces superfluous writes, leading to lower DLWA (92\% less at 10\% zone occupancy), less overall wear (up to 12\%), and up to $3.7\times$ faster workload execution. 
\end{abstract}

\maketitle

\pagestyle{\vldbpagestyle}
\begingroup\small\noindent\raggedright\textbf{PVLDB Reference Format:}\\
\vldbauthors. \vldbtitle. PVLDB, \vldbvolume(\vldbissue): \vldbpages, \vldbyear.\\
\href{https://doi.org/\vldbdoi}{doi:\vldbdoi}
\endgroup
\begingroup
\renewcommand\thefootnote{}\footnote{\noindent
This work is licensed under the Creative Commons BY-NC-ND 4.0 International License. Visit \url{https://creativecommons.org/licenses/by-nc-nd/4.0/} to view a copy of this license. For any use beyond those covered by this license, obtain permission by emailing \href{mailto:info@vldb.org}{info@vldb.org}. Copyright is held by the owner/author(s). Publication rights licensed to the VLDB Endowment. \\
\raggedright Proceedings of the VLDB Endowment, Vol. \vldbvolume, No. \vldbissue\ %
ISSN 2150-8097. \\
\href{https://doi.org/\vldbdoi}{doi:\vldbdoi} \\
}\addtocounter{footnote}{-1}\endgroup

\vspace{0.05in}
\ifdefempty{\vldbavailabilityurl}{}{
\vspace{.3cm}
\begingroup\small\noindent\raggedright\textbf{PVLDB Artifact Availability:}\\
The source code, data, and/or other artifacts have been made available at \url{\vldbavailabilityurl}.
\endgroup
}

\section{Introduction}

\Paragraph{SSDs are Everywhere}
Solid-state drives (SSDs) have become the standard for persistent storage media, offering low-latency access and high bandwidth. Traditional SSDs offer a block-based API so that they can be a drop-in replacement for traditional hard drives, which creates the expectation that performance should be as predictable as it is for hard drives. However, an SSD facing different access patterns, or even SSDs of the same size and with the same underlying flash technology, can exhibit widely varying performance and endurance characteristics, primarily because of their internal design and garbage collection strategies~\cite{Haas2025SSDIq,Lerner2021}.

\Paragraph{Block API in SSDs Causes Garbage Collection}
The block-based device interface provides an abstraction for in-place updates~\cite{Cornwell2012}. 
However, at the physical layer, NAND flash memory has different constraints. 
Flash memory is organized into pages (the unit of read and write) grouped into erase blocks (the unit of erasure).
As a result, NAND flash operates with \emph{out-of-place updates}: once a physical page is written, it cannot be modified until the entire erase block containing it has been erased~\cite{Papon2021a}.
To bridge this semantic mismatch, modern SSDs maintain a flash translation layer (FTL) that maps logical block addresses to physical pages. 
When data is updated, the FTL logically invalidates the old page and writes the new data to a new location, resulting in \textit{out-of-place updates}~\cite{Cornwell2012}. 
The logically invalidated pages are eventually reclaimed through \emph{garbage collection}, during which all valid pages within the partially invalidated erase block are read and rewritten to a new block before the target block is erased.
This leads to significant \emph{device-level write amplification} (DLWA), as more data is written to the physical device than what the host application had requested. 
Over time, this results in performance degradation, wear imbalance, and increased tail latency, particularly under update-heavy workloads~\cite{Bjorling2021ZNS}. 

Prior work on block-based SSDs for data management systems \cite{Athanassoulis2010,Athanassoulis2011,Athanassoulis2014,Stoica2014,Nath2007,Koltsidas2008,Chen2009,Taylor2013,Kang2016,Kang2012,Canim2010,Athanassoulis2014a,Sadoghi2016,Bouganim2009}
was constrained by the limited freedom provided by the FTL. As a result, they primarily focused on avoiding random writes while exploiting efficient random reads with limited control over data placement on the device.

\begin{figure}[!t]
	\includegraphics[width=0.9\columnwidth]{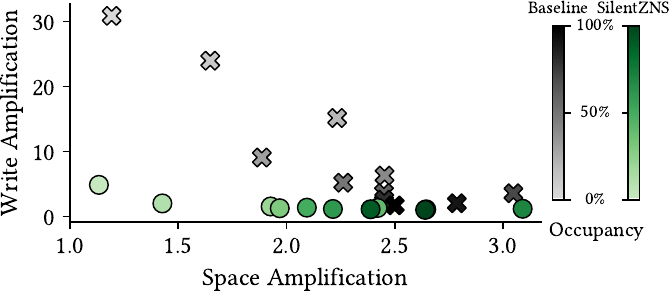}
	\vspace{-2mm}
	\caption{While state-of-the-art ZNS approaches reduce DLWA at the expense of SA, SilentZNS removes the SA overhead. This enables data systems on ZNS reclaim zones according to logical lifetime and free-space needs, rather than delaying reclamation solely to avoid hidden device-side writes.
%
 }
	\label{fig:intro}
	\vspace{-3mm}
\end{figure}

\Paragraph{ZNS: Exposing Flash-Aware Semantics to the Host}
To address the inefficiencies caused by the traditional block interface, Zoned Namespace (ZNS) SSDs introduce a new programming model that aligns more closely with the physical characteristics of flash. 
In ZNS SSDs, the storage is divided into \textit{zones} -- logically contiguous, large regions that must be written sequentially. 
Each zone has a write pointer that advances as data is written, and the entire zone must be erased (via a \texttt{RESET}) before it can be overwritten~\cite{Doekemeijer2023PerformanceCharacterizationON}. 
The sequential-write constraint eliminates the need for complex FTLs and garbage collection, offloading the responsibilities for data placement and I/O scheduling to the host~\cite{Bjorling2021ZNS}. 
This gives the host finer control over data placement and zone state management, thereby shifting complexity from the device to the host/application.
Recent work on ZNS SSDs for data management has focused on LSM-based systems that naturally exhibit immutability \cite{Wu2022ZNSKV,Purandare2022AppendIsN}. In this work, we dissect the internals of ZNS devices to better understand their tradeoffs and usage by data-intensive systems.


\Paragraph{ZNS Operations}
In a ZNS SSD, writes happen via appending a page to a zone, while reading an arbitrary page is supported, similar to a classical SSD.
Additionally, the hosts now manage the zones with two commands: (i) \texttt{RESET} and (ii) \texttt{FINISH}.

\SubParagraph{Zone \texttt{RESET}} When the host decides to reclaim a partially or completely full zone, typically because all written blocks contain invalid data, it issues a zone \texttt{RESET} command. 
This moves the write pointer to the beginning of the zone and marks the previously written blocks as ready for erasure~\cite{Doekemeijer2024ExploringIOM}.

\SubParagraph{Zone \texttt{FINISH}}
Due to limited controller resources, such as memory and power capacitors, ZNS SSDs can support only a limited number of zones (8-32) to be \emph{actively written}~\cite{Bjorling2021ZNS}.
Once this limit is reached, the host has the option to \texttt{RESET} (as discussed above), garbage collect (to another active zone), or to \texttt{FINISH} the zone.
Finishing the zone marks it as read-only and releases the resources needed to write to the recently finished zone back to the controller~\cite{Doekemeijer2023PerformanceCharacterizationON}.

A host manages the zones by explicitly invoking the \texttt{RESET} and \texttt{FINISH} commands to improve performance.
For instance, prior research showed that a LSM-based key-value store running on top of a ZNS SSD can achieve 4$\times$ lower random-read latency and 2$\times$ higher throughput by using the zoned interface compared to block-interface SSD~\cite{Bjorling2021ZNS}.
However, the benefits also depend on how the ZNS standard is implemented by the device's controller~\cite{Long2024WAZone}.

\Paragraph{Challenges with Zone Management}
While the ZNS interface offers benefits, it also introduces new overheads.
When the host issues a \texttt{FINISH} command for a zone that is only partially filled, to satisfy device-level constraints, state-of-the-art ZNS SSDs fill the rest of the zone with dummy data (device-issued writes) before sealing it~\cite{Doekemeijer2024ExploringIOM}. 
This leads to several problems in current ZNS SSDs:

\SubParagraph{1. Unnecessary Write Amplification}
While the dummy writes are invisible to the host, they cause \textbf{significant device-level write amplification (DLWA)}, i.e., a substantial increase in internal writes due to padding, and not due to workload demand~\cite{Doekemeijer2024ExploringIOM}.
Partially used zones that were filled with dummy data must be reclaimed entirely before reuse.
This results in inefficient block utilization and additional erase operations that could be avoided.

\SubParagraph{2. Interference with Host I/O}
Device-issued writes compete with host I/O for the same flash channels and controller resources.
Prior studies show that issuing a \texttt{FINISH} command concurrently with a write-heavy workload can significantly degrade write throughput as it interferes with host-issued I/Os~\cite{Doekemeijer2024ExploringIOM}.

\SubParagraph{3. Host-Side Reclamation Complexity}
To reduce DLWA and interference with host I/O, applications such as RocksDB with ZenFS delay running zone \texttt{FINISH} command, by allowing mixing different lifetime data in the same zone.
However, this delays zone reclamation and increases Space Amplification (SA).
Figure~\ref{fig:intro} illustrates this tradeoff as we vary the \texttt{FINISH} threshold from 0\% to 99\%.
DLWA is computed as the ratio of total writes to host writes, while SA is measured as the ratio of invalidated data to host-inserted data, averaged over the workload.
By delaying \texttt{FINISH} command until 90\% zone occupancy, RocksDB incurs roughly 91\% lower write amplification but at the cost of 69\% higher space amplification compared to running \texttt{FINISH} at 10\% zone occupancy.
Therefore, the current ZNS design forces applications to carefully balance SA and DLWA. 
%

\Paragraph{Implications for Data Management}
Dummy writes consume bandwidth that would otherwise serve application requests, thereby increasing flash wear and introducing latency variability that directly affects operations like WAL appends, buffer-pool flushing, and LSM compaction. Current ZNS-backed systems often compensate by delaying \texttt{FINISH} or by mixing data with different lifetimes within the same zone, which, in turn, increases SA and complicates host-side GC. 
As a result, core data system decisions such as compaction, file placement, and log recycling become tightly coupled with (often undocumented) controller behavior rather than being driven purely by logical efficiency.
Similar in spirit to SSD-iq~\cite{Haas2025SSDIq}, our thesis is that storage behavior hidden behind a standard interface can substantially affect data systems. 
Our approach removes this coupling by making \texttt{FINISH} inexpensive enough that data systems can manage zones according to logical data lifetimes rather than device-specific quirks.

\Paragraph{The Solution: \approach{}}
To address these issues, we propose \approach{}, a zone management strategy that enables the host to reduce DLWA and I/O interference by avoiding unnecessary device writes.
To do this, we first decompose zone management strategies into a design space with four dimensions: 
(i) \textit{zone allocation granularity}, which determines the smallest storage unit that must be finished and reset as a whole (e.g., an individual erase block or a group of blocks); (ii) \textit{zone geometry}, which defines how many blocks compose a zone and across which parallel units (e.g., planes, dies, or chips) they are striped; (iii) \textit{write order}, which determines the order in which writes are distributed to blocks; and (iv) \textit{zone mapping and management}, which dictates how physical zones are constructed (e.g., static vs. dynamic).
We systematically explore this design space 
to identify mappings that eliminate unnecessary DLWA while maintaining high throughput and competitive zone allocation time.
SilentZNS achieves this by releasing unused blocks and reallocating them to new zones, therefore \textit{eliminating the need for device-issued writes} to the entire zone.
Finally, based on our observations, we provide guidelines for data systems to better exploit ZNS SSDs, and we advocate for \textit{bounded} flexibility in future ZNS SSDs, where manufacturers expose a small set of configurations, allowing each application to select the most appropriate one.


\Paragraph{Contributions} Our main contributions are as follows.
\begin{itemize}[leftmargin=10pt]
    \item We identify the causes of increased device-level write amplification, increased wear, and unpredictable performance as suboptimal combinations of zone allocation granularity, zone geometry, write order, and zone mapping and management strategy.
    \item We define a spectrum of mapping strategies where the SSD uses a set of allocation units with different granularity. 
    \item We propose \approach{}, a flexible zone allocation strategy that allocates allocation units to the zone on the fly and issues dummy writes only to the partially written allocation units.
    \item We implement \approach{} in the state-of-the-art ZNS SSD emulator ConfZNS++~\cite{Doekemeijer2024ExploringIOM} and benchmark it by varying the zone geometry and management granularity from block to zone level.
    \item We evaluate \approach{} using synthetic and real-world workloads, including RocksDB with ZenFS, and
    demonstrate how \approach{} reduces DLWA by 95.50\% at 90\% \texttt{FINISH} threshold, while keeping SA of 0.42 and providing 3.7$\times$ faster workload execution, thereby simplifying host-side lifetime management for data systems.
    \item We systematically explore the ZNS design space and provide practical guidelines to help applications and data systems make informed decisions when selecting ZNS SSDs, as well as choosing appropriate configurations in future configurable ZNS devices.
\end{itemize}

\section{Zoned Interface Standard}
\label{sec:ZNS-standard}
We now present the necessary background on the ZNS standard.

\begin{figure}[!t]
	\centering
	\includegraphics[width=1.0\columnwidth]{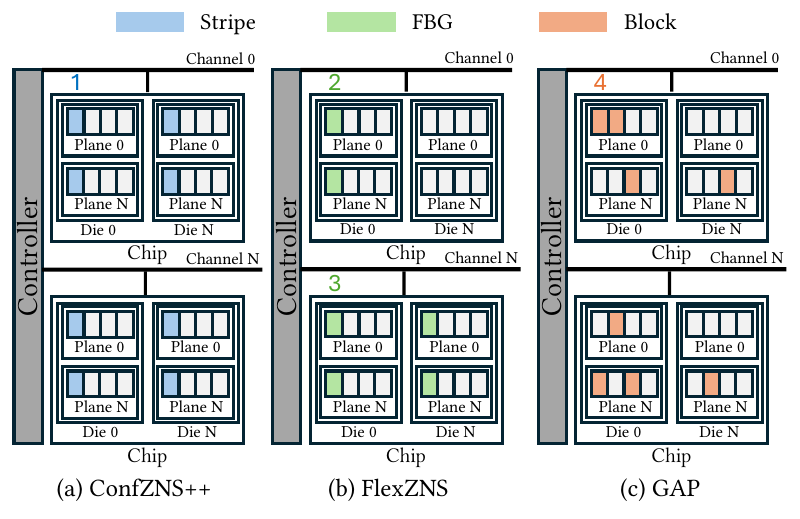}
	\vspace{-6mm}
    \caption{ConfZNS++ constructs zones from multiple superblocks, where each superblock spans across all planes. FlexZNS organizes erase blocks into Flash Block Groups (FBGs) spanning a fixed number of dies. GAP represents chip as a pool of erase blocks and dynamically reconstructs zones from these blocks.}
	\label{fig:background}
	\vspace{-4mm}
\end{figure}

\Paragraphnp{Traditional SSDs} consist of multiple chips connected to a controller via channels \cite{Agrawal2008}.
As shown on Figure~\ref{fig:background}, each chip includes a hierarchy of dies, planes, blocks, and pages, enabling high parallelism~\cite{Cornwell2012,Papon2021a,Papon2021}.
Concurrent I/O is essential to harness the available bandwidth~\cite{Agrawal2008,Papon2023, Papon2024a, Papon2024b}.
Data is written/read at the \emph{page} level but erased at the \emph{block} level. Due to flash’s \emph{erase-before-write} constraint, SSDs perform out-of-place updates, marking old pages invalid. This leads to blocks with both valid and stale data \cite{Stoica2013}, requiring \emph{garbage collection} \cite{Hu2011,Bux2010} to reclaim space, which results in
device-level write amplification, increased write latency, and accelerated wear of flash cells~\cite{Agrawal2008,Papon2021a}.

\Paragraph{ZNS SSDs} To address these limitations, the zoned storage interface organizes logical block addresses (LBAs) into zones, bringing the storage abstraction closer to the organization of the device~\cite{Bjorling2021ZNS}.
Every zone has a write pointer that allows the host to write pages sequentially. 
The ZNS standard also provides the \texttt{APPEND} command, which abstracts away the write pointer by allowing the host to specify only the target zone for the write operation.
Due to hardware constraints, only a limited number of zones can be active at any given time~\cite{Digital2025ZonedStorage}.
The sequential write requirement eliminates device-side garbage collection, as the controller no longer needs to handle out-of-place updates.
Instead, the host assumes the responsibility for both data placement and garbage collection by managing the zone state~\cite{Bjorling2021ZNS},
through two new ZNS commands for zone management: \texttt{RESET} and \texttt{FINISH}. 

\SubParagraph{\texttt{RESET}} When all data in a zone becomes invalid, the host can reclaim it via the \texttt{RESET} command, which moves the write pointer to the zone's beginning~\cite{Doekemeijer2024ExploringIOM}.
The NVMe ZNS standard does not mandate when or how the data must be erased upon a \texttt{RESET}.
Devices choose based on: (i) \texttt{RESET} Timing -- either \emph{synchronous} (erase immediately) or \emph{asynchronous} (invalidate metadata and defer erasure); and (ii) \texttt{RESET} Granularity -- \emph{full} (erase all blocks regardless of content) or \emph{partial} (erase only blocks with invalid data).
While full \texttt{RESET}s are simple to design and maintain even wear across all the erase blocks within the zone, they cause unnecessary wear and delay subsequent writes as full zone must be erased.
Partial \texttt{RESET}s reduce wear and latency as reset only applies to the blocks with invalid data~\cite{Doekemeijer2024ExploringIOM}.

\SubParagraph{\texttt{FINISH}} A zone transitions into \emph{full} state when it is completely filled with data. However, the host may also mark the zone as \emph{full} using the \texttt{FINISH} command, moving the zone's write pointer to the end and preventing any further writes~\cite{Doekemeijer2024ExploringIOM}.
Issuing \texttt{FINISH} also indicates to the controller to release the resources associated with that zone, allowing a new zone to be opened ~\cite{Digital2025ZonedStorage}.
This is useful when the host has reached the limit of \texttt{open} zones but needs to open a new one.
In current designs, when a partially written zone is \emph{finished}, the controller issues dummy writes to fill the remaining space~\cite{Doekemeijer2024ExploringIOM}. 
This is due to flash reliability issues -- leaving pages erased too long can cause program and read disturbs~\cite{Cai2015ReadDisturbEM}. 
Filling up the zone with dummy writes mitigates these risks.
Prior work also proposes using the unwritten pages as a read cache instead of filling them with dummy data.
However, this requires maintaining backup mappings and additional controller-side scheduling logic, increasing controller complexity~\cite{Zhu2023TurnWasteW}.

\Paragraph{Zone Mapping} 
The SSD controller handles logical zone to physical zone mapping: (i) \emph{static mapping} assigns a fixed physical zone to each logical zone, offering control over placement but shifts the burden of wear-leveling to the host~\cite{Doekemeijer2024ExploringIOM}, while (ii) \textit{dynamic mapping} allocates physical zones on demand.
If a free physical zone is available, the controller maps it directly; otherwise, it finds an invalidated (already reset) zone and erases it before reallocating. 
This allows the controller to manage wear and amortize \textit{reset latency}~\cite{Doekemeijer2024ExploringIOM}, an important aspect of zone management, as we also see in our experimentation.

Due to the hierarchical structure of SSDs, physical zones can be (i) mapped to a fixed set of erase blocks or (ii) dynamically constructed.
ConfZNS++ SSD emulator constructs a physical zone from a fixed set of superblocks where a superblock is composed of erase blocks at the same offset across all planes, as demonstrated by allocation unit 1 in Figure~\ref{fig:background}(a)~\cite{Doekemeijer2024ExploringIOM}.
As shown in Figure~\ref{fig:background}(b), FlexZNS organizes storage into Flash Block Groups (FBGs), where each FBG is composed of erase blocks at the same offset across a fixed number of dies~\cite{Wang2023FlexZNS}.
As demonstrated by allocation units 2-3 on Figure~\ref{fig:background}(b), FBG can have a variable size, with the largest FBG being a superblock.
GAP treats chip as a pool of erase blocks and constructs the physical zone dynamically as demonstrated by allocation unit 4 in Figure~\ref{fig:background}(c) ~\cite{Liu2024GAP}.
How physical zones are mapped across the hierarchy of parallel units (chips, dies, planes) introduces different tradeoffs in terms of parallelism, DLWA, and zone allocation overhead, which we analyze in Section~\ref{sec:ZNS-Design}.

\section{ZNS Hidden Write Amplification}
\label{sec:ZNS-Design}

\begin{figure}[t]
	\centering
	\includegraphics[width=1.0\columnwidth]{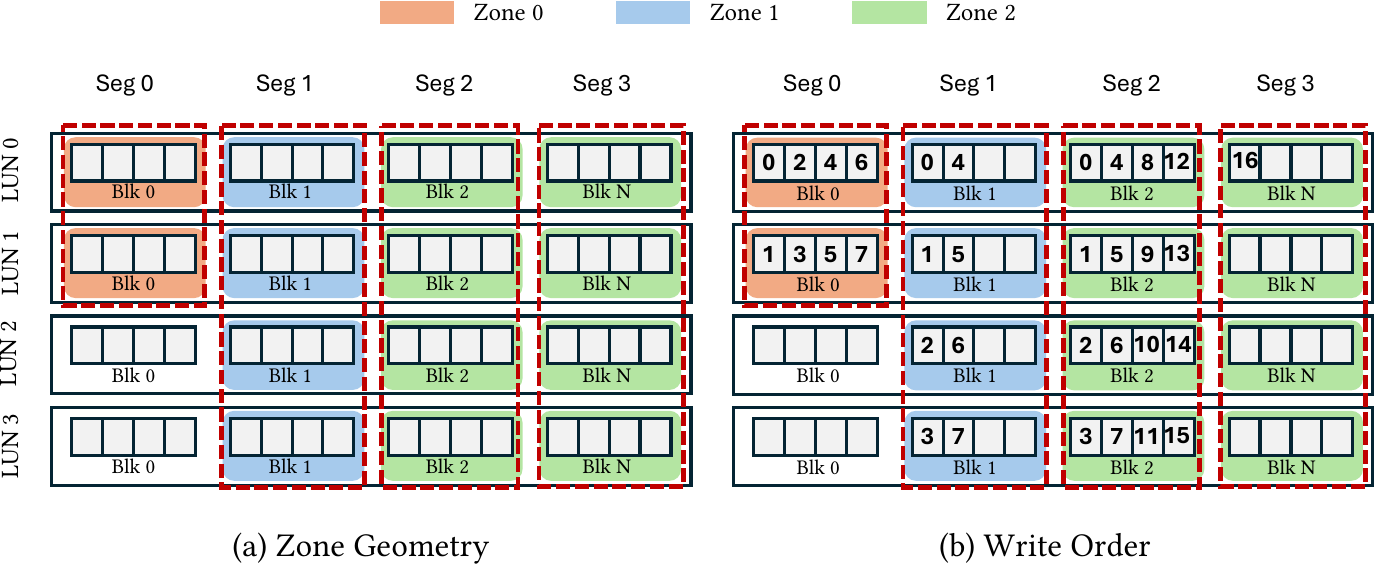}
	\vspace{-6mm}
    \caption{(a) Zone geometry defines the size of a zone and the number of parallel units it spans. (b) Write order controls how data is striped across parallel units.}
	\label{fig:motivation}
	\vspace{-4mm}
\end{figure}

Next, we motivate our work with the hidden write amplification.
We identify four primary design decisions in existing ZNS design space:
(i) \emph{zone geometry}, (ii) \emph{write order}, (iii) \emph{physical zone allocation} and (iv) \emph{zone reclamation}.
For simplicity, throughout the remainder of the paper, we abstract the underlying storage hierarchy and refer to each parallel component of the SSD as a logical unit (LUN), where LUN consists of erase blocks (blk), as shown in Figure~\ref{fig:motivation}.
A zone is constructed from segments, where the segment consists of erase blocks across LUNs.
For example as shown on Figure~\ref{fig:motivation} segment \textit{Seg0} has parallelism of 2 as it spans 2 LUNs.

\SubParagraphno{Zone Geometry} determines the type of segments used in the zone (i.e., parallelism of a segment) and how many segments are used to construct a zone (i.e., its size).
Parallelism of the segment determines the intra-zone parallelism.
For example, \textit{Zone0} in Figure~\ref{fig:motivation}(a) consists of one segment \textit{Seg0} that has a parallelism of 2 therefore intra-zone parallelism is also 2.
\textit{Zone1} in Figure~\ref{fig:motivation}(a) illustrates a zone that maps to segment \textit{Seg1} with parallelism of 4, therefore intra-zone parallelism is 4.
Zones that span all parallel units are designed to maximize intra-zone parallelism~\cite{Doekemeijer2024ExploringIOM}.
In addition to parallel units, prior approaches also vary the zone size.
For example, large zones may also consist of more than one segment as illustrated by \textit{Zone2} in Figure~\ref{fig:motivation}(a).
The benefit of larger zones is less metadata at the FTL level.
However, the downside is that more blocks need to be dummy-written during \texttt{FINISH}, thus increasing DLWA and interference with host I/O.

\SubParagraph{Write Order}
To exploit internal parallelism, prior work stripes data across all parallel units~\cite{Doekemeijer2024ExploringIOM}.
As shown in Figure~\ref{fig:motivation}(b), in zones 0 and 1, pages 0–7 are written in a striped order across the LUNs.
If a zone contains multiple segments, as in \textit{zone 2} of Figure~\ref{fig:motivation}(b), each segment is written to completion before the next segment begins.
For example, segment \textit{Seg2} is fully written, covering pages 0–15.
The write pointer then advances to page 16 and starts writing the next segment \textit{Seg3}.
Striping across all parallel units is designed to take advantage of SSD parallelism and increase intra-zone throughput.
However, this design also causes writes to touch many erase blocks.
During \texttt{FINISH}, all the partially written blocks must be padded with dummy data.
As a result, maximizing intra-zone parallelism can inadvertently increase DLWA.

\SubParagraph{Physical Zone Allocation}
Prior work treats a physical zone as a fixed set of erase blocks that are mapped to logical zones~\cite{Doekemeijer2024ExploringIOM}.  
The advantage of static physical zones is their low allocation latency and reduced metadata overhead, since logical zones can be directly mapped to physical zones.  
Some prior approaches allow the dynamic allocation of erase blocks to zones~\cite{Liu2024GAP}.
This enables more fine-grained control over zone allocation, but it also requires additional metadata to maintain the mapping table between logical zones and allocated blocks.

\SubParagraph{Zone Reclamation}
Prior work proposes full zone reclamation, where if a zone is partially written and the host issues a zone \texttt{FINISH} command, the entire zone is filled with dummy data and consequently the entire zone is erased during \texttt{RESET}.  
This is tied to the fixed physical zone allocation strategy, where erase blocks from one physical zone cannot be assigned to another zone.  
In addition, hardware constraints prevent blocks from remaining in the erased state.  
As a result, the entire zone must be filled with dummy data and subsequently erased during reset, causing unnecessary DLWA~\cite{Doekemeijer2024ExploringIOM}.

\begin{figure}[t]
  \centering
  \includegraphics[width=\columnwidth]{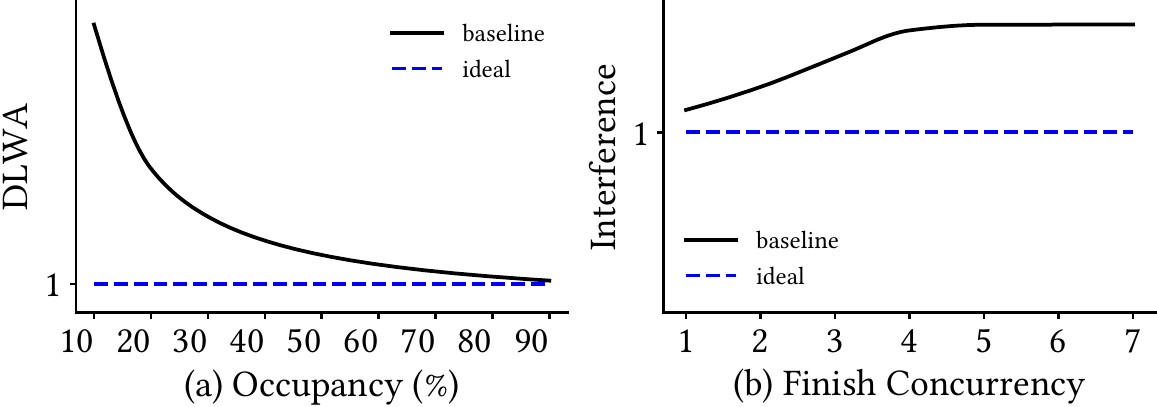}
  \vspace{-6mm}
  \caption{(a) DLWA decreases as zone occupancy increases, approaching the ideal baseline at high occupancy. (b) Concurrent \texttt{FINISH} operations interfere with host writes.}
  \vspace{-4mm}
  \label{fig:motivation-exp}
\end{figure}

\Paragraph{Hidden Write Amplification in State-of-the-art Devices}
Now we demonstrate how these design decisions can cause write amplification using ConfZNS++~\cite{Doekemeijer2024ExploringIOM}, a state-of-the-art ZNS SSD emulator that emulates a real ZNS SSD (ZN540).
The emulated SSD uses fixed physical zones, follows the large-zone model, and stripes writes across all parallel units~\cite{Doekemeijer2024ExploringIOM}.
While this design is optimized for intra-zone parallelism and low physical zone allocation time, it increases dummy writes and, thus, DLWA.
Figure~\ref{fig:motivation-exp}(a) shows that DLWA is significantly higher at low occupancy because finishing a partially written zone forces dummy writes across all erase blocks in the zone.
Figure~\ref{fig:motivation-exp}(b) shows the interference caused by concurrent \texttt{FINISH} operations. 
For example, a \texttt{FINISH} concurrency level of 1 means that one zone is being finished while another zone is written by the host. 
We define the interference factor as the ratio of baseline throughput to throughput under concurrent \texttt{FINISH} operations.
Higher values indicate greater interference and, thus, lower throughput. 
Figure~\ref{fig:motivation-exp}(b) shows that, as we increase the number of zones being concurrently finished, \texttt{FINISH} operations interfere with host writes, reducing throughput.
Therefore, static zone management designs that maximize intra-zone parallelism through full striping to reduce allocation overhead increase DLWA and I/O interference.
We present more details in Section~\ref{sec:ZNS-Evaluation}.

\section{Augmenting the ZNS Design Space}
\label{sec:ZNS-Design-space}

\begin{figure*}[t]
    \centering
    \includegraphics[width=\linewidth]{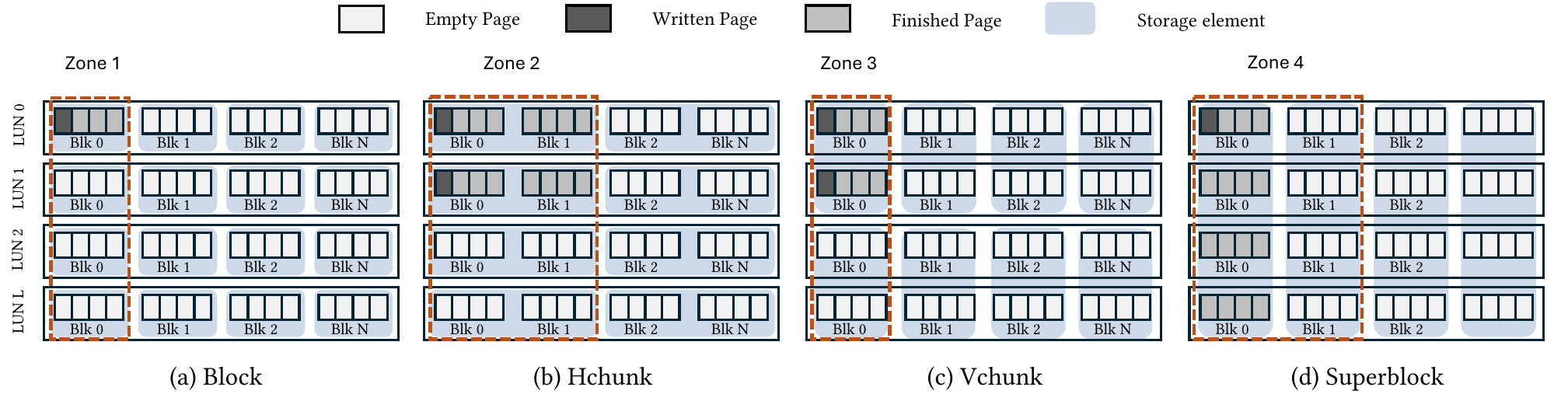}
    \vspace{-6mm}
    \caption{(a) Zone 1 is constructed from $L$ storage elements, where each storage element is an erase block. 
    (b) Zone 2 is constructed from four storage elements, which are horizontal chunks (Hchunks), where each Hchunk contains two erase blocks. 
    (c) Zone 3 is constructed from two storage elements, which are vertical chunks (Vchunks), where each Vchunk contains two erase blocks. 
    (d) Zone 4 is constructed from a superblock, which consists of one erase block from each LUN.}
    \label{fig:alloc_strategies}
    \vspace{-2mm}
\end{figure*}

We now present an augmented design space for allocating physical blocks to zones, which serves as a tool to help us unlock better zone allocation strategies. Table~\ref{tab:storage_terms} summarizes the terms we use.

\setlength{\belowcaptionskip}{0pt}  
\setlength{\textfloatsep}{2pt}
\begin{table}[h]
{
\begin{center}
\caption{Terms to describe the augmented ZNS design space}
\vspace{-2mm}
\begin{tabular}{cl}
\toprule
\textbf{Term} & \textbf{Description} \\
\midrule
Block & An SSD erase block \\
LUN & Parallel unit consisting of multiple blocks \\
Superblock & Consists of one block per LUN \\
Vchunk-X & Group of X blocks across adjacent LUNs \\
Hchunk-X & Sequence of X blocks within a LUN \\
Storage element & Basic unit used to construct zones \\
Zone segment & Maximum subset of a zone \\
                & that fully exploits zone's LUN parallelism \\
\bottomrule
\end{tabular}
\label{tab:storage_terms}
\end{center}
}
\vspace{-4mm}
\end{table}

\Paragraph{The Augmented ZNS Design Space}
Constructing the optimal physical zone requires balancing four key metrics: throughput, DLWA, allocation time, and wear-leveling.
To balance these metrics, we augment the \textit{write order} design dimension and \textit{zone reclamation strategy} and introduce a new dimension \textit{zone allocation granularity}. 

\Paragraph{\textit{Augmented Dimension:} Write Order}
The write order that fully stripes pages across all parallel units may inadvertently increase DLWA and interfere with host I/O.
We consider an alternative write-order strategy that stripes writes across a fixed subset of parallel units within a zone.
This design introduces a tradeoff between throughput and DLWA.
Striping across fewer parallel units reduces the number of partially written erase blocks that must be written with dummy data during \texttt{FINISH}, therefore lowering DLWA and mitigating interference with host I/O.
This can be simulated by adjusting the zone geometry configuration.
For example, smaller zones such as \textit{Zone0} in Figure~\ref{fig:motivation}(b) would enforce striping data across fewer parallel units compared to \textit{Zone1} that stripes data across all parallel units.
However, limiting parallelism may reduce the intra-zone throughput.
Therefore, the degree of parallelism becomes a tunable parameter that balances throughput and DLWA.

\Paragraph{\textit{Augmented Dimension:} Zone Reclamation}
Prior designs fills the remaining empty blocks of a zone with dummy data regardless of the occupancy of a zone, causing DLWA. 
We introduce a new \texttt{FINISH} design in which only the minimum required erase blocks are written with dummy data and subsequently erased during the \texttt{RESET} operation. 
The remaining empty blocks are released and available for allocation to the next zone.
This is possible by allowing physical zones to be allocated dynamically.

\Paragraph{\textit{New Dimension:} Zone Allocation Granularity}
The new zone reclamation strategy introduces an additional dimension in the design space: the zone allocation granularity.
Specifically, it defines the smallest physical unit from which a zone is constructed. 
Choosing the size and geometry of this unit involves balancing throughput, DLWA, allocation time, and wear-leveling, since each unit is finished and reset as a whole.
We propose an augmented design space in which storage is represented as a collection of $L$ parallel units (LUNs), each consisting of $N$ erase blocks, as illustrated in Figure~\ref{fig:alloc_strategies}.
Zone allocation occurs using five different types of storage elements: (i) block, (ii) horizontal chunk, (iii) vertical chunk, (iv) superblock, (v) fixed physical zone.
Next, we discuss the advantages and drawbacks of each option.

\SubParagraph{Block}
Prior work proposes the allocation of zones from a pool of erase blocks~\cite{Liu2024GAP}.
However, they do not design or evaluate the impact of the \texttt{FINISH} command.
In the augmented design, we model storage as a collection of erase blocks and construct each zone such that blocks are evenly distributed across parallel units to ensure the intra-zone parallelism as illustrated in Figure~\ref{fig:alloc_strategies}(a).
In section 5, we describe the constraints required to preserve intra-zone parallelism in more detail.
The primary advantage of block-based allocation is that is allows more fine-grained control over DLWA.
During a \texttt{FINISH} operation, only the partially written erase blocks need to be written with dummy data.
For example, Figure~\ref{fig:alloc_strategies}(a) shows that only partially written block \textit{Blk0} from \textit{LUN0} is written with dummy data while the rest of the empty blocks can be released and allocated to the next zone.
The downside is higher zone allocation latency due to searching through erase blocks to construct the zone, along with additional metadata overhead to maintain the mapping between logical zones and allocated erase blocks.

\SubParagraph{Horizontal Chunk}
To reduce the search space and required metadata, we introduce horizontal chunks (Hchunk).
As shown in Figure~\ref{fig:alloc_strategies}(b), a horizontal chunk is a contiguous sequence of erase blocks within a single parallel unit.
To ensure parallelism, a zone is constructed from chunks across a set number of parallel units.
The main advantage is a smaller search space, which reduces the zone allocation time.
However, as illustrated in Figure~\ref{fig:alloc_strategies}(b), the limitation is that if any page within a chunk is written, all remaining erase blocks in that chunk must be padded with dummy data during \texttt{FINISH}, even if they were never used.
Therefore, increasing the chunk size may reduce the zone allocation and metadata overhead, but may increase DLWA and interference with host I/O.

\SubParagraph{Vertical Chunk}
Vertical chunks group erase blocks across adjacent LUNs, illustrated in Figure~\ref{fig:alloc_strategies}(c).
The maximum vertical chunk size corresponds to a full stripe of erase blocks spanning all LUNs.
As with horizontal chunks, during the zone \texttt{FINISH} operation, all partially written vertical chunks must be filled with dummy data.
Therefore, as the chunk size increases, both DLWA and interference with host I/O increase.
However, compared to horizontal chunks, vertical chunks have an important advantage: they align better with striped writes across parallel units.
For example, in Figure~\ref{fig:alloc_strategies}, both Hchunk and Vchunk have the same size (two erase blocks), and the writes are striped across parallel units.
In this case, Hchunk requires padding four erase blocks with dummy data, whereas Vchunk requires padding only two, due to the geometry of the storage element.

\SubParagraph{Superblock}
A vertical chunk that spans all parallel units is called a superblock, as shown in Figure~\ref{fig:alloc_strategies}(d). 
A superblock consists of one erase block from each parallel unit. 
Using a superblock as the allocation unit simplifies zone construction, as there is no need to explicitly enforce parallelism constraints. 
The superblock inherently ensures that all parallel units contribute one block. 
However, as illustrated in Figure~\ref{fig:alloc_strategies}(d), the downside is that the superblock has more coarse-grained allocation granularity and during a \texttt{FINISH} operation, all erase blocks in the superblock must be padded with dummy data, even if some were never written.

\SubParagraph{Fixed physical zone}
In this mapping scheme, the physical zone is mapped to a fixed set of erase blocks.
The advantage of this direct mapping is that allocation time and metadata size are kept minimal.
The downside is that due to the fixed physical zone design, dynamic zone reclamation is not possible.
Therefore, all empty blocks in the zone must be filled with dummy data during \texttt{FINISH}, causing interference with host I/O and unnecessary wear, as these blocks must later be erased once they are allocated to a new logical zone.


\section{\texttt{S\lowercase{ilent}ZNS}: Flexible Zone Allocation}

\Paragraph{Design Goals}
We design \approach{}, a fine-grained device-level zone management strategy for ZNS SSDs that represents storage as a collection of \textit{storage elements} with configurable sizes and allocates the storage elements to logical zones on demand. 
By configuring the granularity of the storage element, \approach{} navigates the continuum of ZNS design space -- from fine-grained block-level allocation to coarse-grained superblock allocation. 
\approach{} provides a flexible abstraction for zone allocation, introducing tradeoffs in allocation latency and zone management granularity. 

\Paragraph{From Blocks to Superblocks: The Storage Continuum}
The flexible deisgn of \approach{} allows it to model a continuum of ZNS zone allocation strategies -- from a \textit{block}-based allocation that optimizes for device write amplification to a \textit{superblock}-based allocation that optimizes for zone allocation time.  
\approach{} allocates zones based on two metadata: (i) \textit{wear} -- the number of times a storage element has been erased, and (ii) \textit{availability} 
\( a \), indicating the element's status: \( a = 0 \) (empty and available for allocation), \( a = 1 \) (allocated but still empty), \( a = 2 \) (allocated and contains valid data) and \( a = 3 \) (free for (re-)allocation but contains invalid data).


\Paragraph{\approach{} with Block and Hchunk Allocation}
Block-based zone allocation in \approach{} aims to minimize DLWA, while Hchunk aims to reduce zone allocation time.
In this formulation, we consider block allocation as a special case of Hchunk allocation, where $c_s$ denotes the chunk size and $c_s=1$. 
Therefore, we use Hchunk to describe the formulation for the remainder of this subsection. Our full notation is summarized in Table~\ref{tab:rl_terms}.
A \textit{Hchunk} is treated as the smallest storage element, and zones are built on the fly using one or more Hchunks per LUN. 
The Hchunk allocation algorithm must (i) support wear-leveling and (ii) maintain parallelism by evenly distributing Hchunks across LUNs.
Hence, the objective of the Hchunk allocation algorithm is to minimize wear by selecting Hchunks with the lowest wear and to select $G$ Hchunks from each LUN.
We formulate the objective as:
\[
\text{Minimize} \quad \sum\nolimits_{n=1}^{N} c_n \cdot w_n
\]
where \( c_n \in \{0,1\} \) indicates if Hchunk \( n \) is selected (\( c_n = 1 \)) and 
\( w_n \) is the wear associated with Hchunk \( n \). 
A chunk must satisfy the following constraints to ensure that we use Hchunks with the lowest wear while maintaining parallelism: (i) \emph{chunk availability}, (ii) \emph{chunk selection}, and (iii) \emph{zone parallelism}.

\setlength{\belowcaptionskip}{-6pt}  
\setlength{\textfloatsep}{2pt}
\begin{table}[t]
{
\begin{center}
\small
\caption{Notation for modeling ZNS SSD}
\vspace{-4mm}
\begin{tabular}{cl}
\toprule
\textbf{Term} & \textbf{Definition} \\
\midrule
$N$ & total number of storage elements \\
$L$ & total number of LUNs \\
$Z$ & number of storage elements selected for a zone \\
$K$ & max number of selected chunks from an active LUN \\
$w$ & storage element wear array of size $N$ \\
$a$ & storage element availability array of size $N$ \\
$c$ & chunk selection array of size $N$, where $c_n \in \{0,1\}$ \\
$c_s$ & chunk of size $s$ erase blocks \\
$s$ & LUN activation array of size $L$, where $s_l \in \{0,1\}$ \\
$g_l$ & chunk membership vector for LUN $l$ \\
$L_{\min}$ & minimum number of active LUNs required for a zone \\
$L_{\mathrm{elig}}$ & set of LUNs eligible for allocation \\
\bottomrule
\end{tabular}
\label{tab:rl_terms}
\end{center}
}
\end{table}

\SubParagraphno{Hchunk availability}
The availability constraint ensures that only available (i.e., $a_n \in \{0,3\}$) chunks can be selected.
In the implementation, unavailable chunks are excluded by fixing their decision variables to zero:
\begin{align}
    c_n = 0 \quad \text{if } a_n \notin \{0,3\}, \quad \forall n \in \{1, \ldots, N\}
\end{align}

\SubParagraphno{Hchunk selection}
This constraint ensures that each zone is allocated a fixed number of chunks $Z$.
The ZNS standard requires each zone to have equal size; therefore, the total number of selected chunks must equal $Z$:
\begin{align}
    \sum\nolimits_{n=1}^{N} c_n = Z
\end{align}

\SubParagraphno{Zone parallelism}
To ensure that the zone meets parallelism requirements, we introduce binary variables \( s_l \in \{0,1\} \), which indicate whether LUN \( l \) participates in constructing the zone.
\( s_l = 1 \) means that LUN is active and \( l \) contributes at least one chunk, while \( s_l = 0 \) means it is not active.
To ensure sufficient parallelism, at least \(L_{\min}\) LUNs must contribute to zone:
\begin{align}
    \sum\nolimits_{l=1}^{L} s_l \geq L_{\min}
\end{align}

To model chunk-to-LUN relationships, we define \( g_l \in \{0,1\}^N \) as the membership of chunks in LUN \( l \), where \( g_l[n] = 1 \) if chunk \( n \) belongs to LUN \( l \).
The total number of selected chunks from LUN \( l \) is given by \( \sum_{n=1}^{N} (c_n \cdot g_l[n]) \).
We couple chunk selection with LUN participation using the following constraints:
\begin{align}
    \sum\nolimits_{n=1}^{N} (c_n \cdot g_l[n]) \geq s_l \quad \forall l \in \{1, \ldots, L\}
\end{align}
\begin{align}
    \sum\nolimits_{n=1}^{N} (c_n \cdot g_l[n]) \leq K \cdot s_l \quad \forall l \in \{1, \ldots, L\}
\end{align}

Constraint (5) ensures that if a LUN is marked as active, it must contribute at least one chunk.
Constraint (6) bounds the number of chunks assigned to each LUN.
If \( s_l = 0 \), the upper bound forces the number of selected chunks to be zero, preventing allocation from inactive LUNs.
If \( s_l = 1 \), the LUN can contribute up to \( K \) chunks.
Here, \( K \) denotes the maximum number of chunks that can be selected from an active LUN.
With these constraints, the Hchunk-based mapping strategy enables flexible allocation across LUNs while maintaining bounded parallelism.
By tuning \(L_{\min}\) and \(K\), the allocator can relax parallelism requirements in favor of improved wear-leveling.

Additionally, since we model zone geometries with varying levels of parallelism, when the desired zone parallelism is smaller than the total number of LUNs \(L\), we select LUNs in a round-robin manner for each zone allocation.
This approach reduces inter-zone interference across consecutive allocations.
If zones were allocated on the same set of LUNs, they would compete for the same write resources during host-issued writes.
Thus, for each allocation, we identify a set of eligible LUNs, and LUNs outside the eligible set \( \mathcal{L}_{\mathrm{elig}} \) are disabled:
\begin{align}
    s_l = 0 \quad \forall l \notin L_{\mathrm{elig}}
\end{align}


\Paragraph{\approach{} with Vchunk Allocation}
Vchunk-based zone allocation in \approach{} aligns allocation with striped write order while reducing metadata overhead and allocation time.
A Vchunk groups storage elements across \(c_s\) consecutive LUNs, where \(1 \leq c_s \leq L\) and \(c_s\) must evenly divide \(L\).
When \(c_s = 2\), each Vchunk consists of 2 blocks that span 2 LUNs.
Vchunk allocation does not change the physical number of LUNs.
Instead, it just groups \(c_s\) LUNs into a single logical group.
Thus, the allocator operates over \(L_{\text{grp}} = L / c_s\) logical groups.
Therefore, Vchunk allocation uses the same formulation as Hchunk allocation, with the only change being the definition of the storage element.
Each decision variable \(c_n\) now corresponds to selecting a Vchunk, and the allocator operates over \(L_{\text{grp}}\) logical groups instead of L LUNs.

In Vchunk allocation by varying \(c_s\) we can vary the Vchunk size.
Smaller Vchunks provide finer-grained control over DLWA but at the cost of higher metadata overhead.
Larger Vchunks reduce metadata and allocation time but increase DLWA, as more unwritten pages within a Vchunk must be filled with dummy data.
At the extreme, when \(c_s = L\), each Vchunk spans all LUNs and becomes a superblock, providing maximum intra-zone parallelism but requiring full padding during zone \texttt{FINISH}.

\Paragraph{\approach{} with Superblock Allocation}
Superblock-based zone allocation in \approach{} is optimized for low zone allocation time and less metadata.
A \textit{superblock} is a fixed set of erase blocks across all LUNs and is treated as the smallest storage element.
As shown in Figure~\ref{fig:alloc_strategies}d, superblock-based allocation constructs zones from superblocks, each with one block per LUN. 
Superblock abstraction inherently enforces parallelism across all LUNs, removing the need for an explicit parallelism constraint and, therefore, simplifying the search space.
The objective of the superblock allocation algorithm is to minimize wear by selecting superblocks with the lowest wear.
We formulate the objective as follows:
\[
\text{Minimize} \quad \sum\nolimits_{n=1}^{N} s_n \cdot w_n
\]
where,  \( s_n \in \{0,1\} \) indicates whether superblock \( n \) is selected and \( w_n \) is the wear associated with superblock \( n \).
The selected superblocks must satisfy the \textit{superblock availability} and \textit{superblock selection} constraints:

\SubParagraphno{Superblock availability} ensures that only available
superblocks (\( a_n = 0 \) and \( a_n = 3 \)) can be selected for the zone.
\begin{align}
    s_n \leq [a_n = 0 \lor a_n = 3] \quad \forall n \in \{1, \ldots, N\}
\end{align}

\SubParagraph{Superblock selection} This constraint ensures that each zone is allocated exactly $Z$ superblocks.
Since every zone must maintain a fixed size, the total number of selected superblocks is equal to $Z$, i.e.,
\begin{align}
    \sum\nolimits_{n=1}^{N} s_n = Z
\end{align}
The main advantage of superblock-based allocation over chunk-based allocation is that it inherently satisfies parallelism constraints due to the definition of superblocks.  
However, the trade-off is reduced flexibility.  
Because a superblock spans across all LUNs and is treated as a single storage element, even a single page write requires the entire superblock to be written with dummy data during the \texttt{FINISH} command.  
As a result, \texttt{FINISH} operates at a coarser granularity.

\begin{figure*}[!t]
    \centering
    \includegraphics[width=\textwidth]{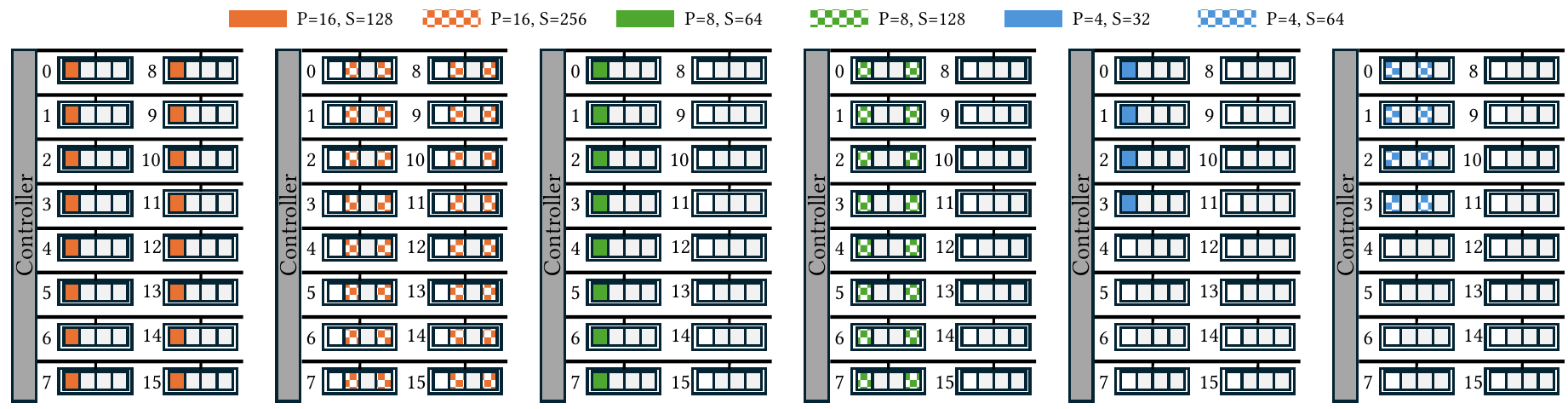}
    \caption{Illustration of the six zone geometry configurations with varying zone parallelism P and zone size S.}
    \label{fig:geometry-configurations}
\end{figure*}

\Paragraph{Integration with SSD Emulator}
We use MOSEK to implement \approach{} and integrate it into the \texttt{Zone Allocator}, \texttt{READ}, \texttt{WRITE}, \texttt{FINISH}, and \texttt{RESET} modules in the \texttt{ConfZNS++} SSD emulator.

\SubParagraph{\texttt{Zone Allocator}}
The allocator maintains a mapping table between zones and their storage elements. When a zone receives its first write, \approach{} allocates storage elements, marks them as allocated (\( a = 1 \)), and adds them to the table.

\SubParagraph{\texttt{READ/WRITE}}
Once \approach{} receives read/write request, \approach{} computes the zone index from the LBA and retrieves its storage elements from the table. LBAs are always superblockd sequentially across elements. For writes, the metadata is updated to valid (\( a = 2 \)).

\SubParagraph{\texttt{FINISH}}
With on-demand allocation, \texttt{FINISH} is applied at the storage element level (Figure~\ref{fig:alloc_strategies}). 
Dummy writes are issued only to partially written elements to meet hardware constraints (Section~\ref{sec:ZNS-Design}).
After \texttt{FINISH}, \approach{} releases unused storage elements (\( a = 1 \)) from the mapping table and marks them free  (\( a = 0 \)), making them available for future zone allocations, while written ones (\( a = 2 \)) remain mapped to support reads.
Though DLWA is reduced, some dummy writes may still occur based on mapping granularity.
 

\SubParagraph{\texttt{RESET}}
We support partial, asynchronous \texttt{RESET} as in ConfZNS++ and ZN540.
On \texttt{RESET}, written elements (\( a = 2 \)) are marked invalid (\( a = 3 \)), indicating that they need to be erased first before reuse.
As for empty storage elements (\( a = 1 \)), they are marked as free (\( a = 0 \)) so they can be reused.
After updating the metadata, \approach{} releases these storage elements from the mapping table.
Once they are mapped to another logical zone, they are erased physically.

\section{Evaluation}
\label{sec:ZNS-Evaluation}
We now demonstrate the benefits of flexible zone management. 

\subsection{Evaluation Methodology}

\Paragraph{SSD Emulator}
The SSD emulator used in this work, ConfZNS++, is built on top of FEMU.
FEMU enables fine-grained control over internal device parameters such as channels, chips, dies, and planes, as well as configurable page-level read, program, and erase latencies.
From the host’s perspective, a FEMU-based device behaves like a standard NVMe SSD, allowing applications to interact with the emulated device as if it were real hardware~\cite{Li2018FEMU}.

\Paragraph{Emulated SSD Configuration} Throughout this paper, we use two SSD configurations: (1) a model of a real ZNS SSD (ZN540) and (2) a custom SSD design to explore a broader range of zone geometries and allocation strategies.

\SubParagraph{Baseline ZNS SSD}
We use the ConfZNS++ emulator to model a Western Digital ZN540 ZNS SSD. The device has 4 channels, a page size of 16~KiB, and a block size of 768 pages.
It provides a zone capacity of 1~GiB, with a total of 48 zones and a limit of 14 open and active zones.
The SSD supports intra-zone parallelism of 4 and simulates latencies of 700~\textmu s for writes, 60~\textmu s for reads, and 3.5~ms for erases~\cite{Doekemeijer2024ExploringIOM}.
It follows a fixed physical zone design where physical zone consists of 22 superblocks where each superblock contains 4 erase blocks.


\SubParagraph{Custom SSD Configuration}
To explore different zone geometries and allocation strategies, we also use a custom SSD model.
This SSD has 8 channels and 2 ways, resulting in a total of 16 LUNs.
It uses a page size of 4~KB and a block size of 2048 pages. The device simulates latencies of 500~$\mu$s for writes, 50~$\mu$s for reads, 25~$\mu$s for channel transfer, and 5~ms for erases~\cite{Wang2023FlexZNS}.
The total number of zones and zone capacity depend on the zone geometry configuration.

\Paragraph{Zone Geometry Configurations}
Using our custom SSD configuration, we evaluate six different zone geometry configurations. Figure~\ref{fig:geometry-configurations} illustrates these configurations. Each box, labeled 0--15, represents a LUN consisting of multiple erase blocks.
P denotes the zone parallelism which refers to how many LUNs contribute to the zone.
S denotes the zone size, measured in MiB.
A configuration with $P=16$ and $S=128$~MiB represents a zone that is constructed from one zone segment that is a superblock (128~MiB).
A configuration with $P=16$ and $S=256$~MiB increases the zone size by using two segments where each segment is a superblock.
A configuration with $P=8$ and $S=64$~MiB models a smaller zone that uses segments of parallelism 8 (64~MiB).
A configuration with $P=8$ and $S=128$~MiB increases the zone size to $S=128$~MiB by using two segments of parallelism 8.
Finally, the configuration with $P=4$ and $S=32$~MiB uses one segment of parallelism 4 (32~MiB), while $P=4$ and $S=64$~MiB increases zone size by using two segments.

\begin{figure*}[t]
  \centering
  \begin{subfigure}{.24\textwidth}
    \centering
    \includegraphics[width=\linewidth]{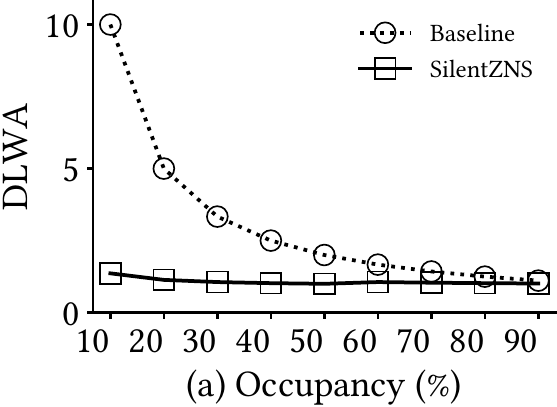}
    \label{fig:dlwa}
  \end{subfigure}
  \begin{subfigure}{.24\linewidth}
    \centering
    \includegraphics[width=\linewidth]{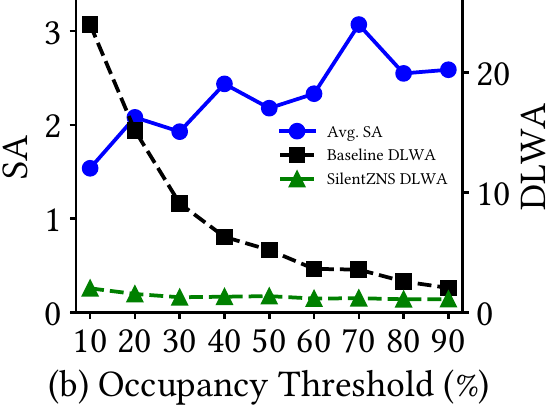}
    \label{fig:host-garbage}
  \end{subfigure}
  \begin{subfigure}{.24\textwidth}
    \centering
    \includegraphics[width=\linewidth]{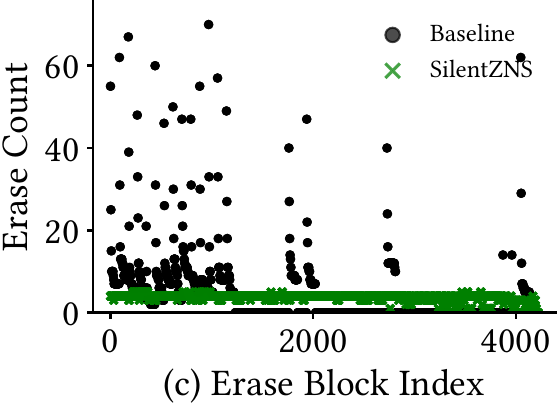}
    \label{fig:wear}
  \end{subfigure}
  \begin{subfigure}{.24\linewidth}
    \centering
    \includegraphics[width=\linewidth]{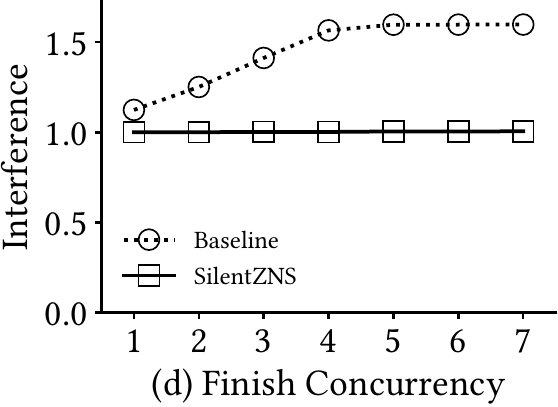}
    \label{fig:device-garbage}
  \end{subfigure}
  \vspace{-6mm}
  \caption{(a) \approach{}'s DLWA benefits are highest at low zone occupancy. (b) A Lower occupancy threshold reduces SA. (c) \approach{} distributes wear more evenly. (d) \approach{} reduces interference with host writes. }
  \label{fig:raw_bench}
  \vspace{-2mm}
\end{figure*}

\Paragraph{Zone Storage Elements}
With our custom SSD configuration, we evaluate six zone allocation granularities: \textit{fixed}, \textit{superblock}, \textit{block}, \textit{Vchunk-2}, \textit{Vchunk-4}, and \textit{Hchunk-2}.
The \textit{fixed} allocation corresponds to the baseline used in ConfZNS++, where physical zone is mapped to a fixed set of erase blocks.
A \textit{superblock} consists of one block from each of the 16 LUNs, resulting in a size of 128~MiB.
In total, there are 128 superblocks. The \textit{block} allocation uses a single erase block as the allocation unit.
For vertical chunks, the chunk size must evenly divide the total number of LUNs (16). A Vchunk of size 16 is equivalent to a superblock. In our evaluation, we also consider finer-grained vertical chunks with sizes 2 and 4, referred to as \textit{Vchunk-2} and \textit{Vchunk-4}, respectively.
The \textit{Hchunk-2} allocation groups two consecutive blocks within a LUN to form a horizontal chunk.
Note that once all blocks within a segment are written under \textit{Hchunk-2} allocation, it behaves equivalently to \textit{fixed} allocation in terms of the zone reclamation.
This highlights that \approach{} exposes a continuum of storage elements, allowing our formulation to represent a wide range of configurations.

\Paragraph{Workloads}
We evaluate using custom benchmarks, FIO~\cite{Fio} and KVBench~\cite{Zhu2024a}. 
Our workloads are:

\SubParagraph{DLWA Benchmark}
We evaluate the cost of \texttt{FINISH} operations by filling zones to a target occupancy level.
Then issue \texttt{FINISH} and record the number of pages finished.

\SubParagraph{Interference Benchmark}
We measure the impact of concurrent \texttt{FINISH} on host I/O.
A \texttt{FINISH} concurrency level of $N$ indicates that $N$ partially filled zones are being finished while the host writes to N other zones concurrently.

\SubParagraph{Raw-device Write Benchmark}
FIO-based sequential write workload where we vary the request size within a zone and the number of concurrently written zones.
Unless otherwise stated, FIO jobs use requests with direct, synchronous I/O, writing to a dedicated zone.

\SubParagraph{KVBench-II}
This mixed workload interleaves 50\% inserts, 10\% deletes, 15\% point queries, and 25\% updates~\cite{Zhu2024a}.
We run KVBench on RocksDB using ZenFS as the filesystem backend.

\Paragraph{RocksDB with ZenFS} 
ZenFS integrates with RocksDB using a zone-based interface and assigns files based on write lifetime hints. If no matching zone is available, it allocates a new one. 
When open/active zone limits are hit, ZenFS issues a \texttt{FINISH}, based on a configurable \texttt{FINISH} threshold (e.g., 10\% remaining space).
Throughout the rest of the paper we express \texttt{FINISH} threshold in terms of occupancy.
A \texttt{FINISH} threshold of 10\% corresponds to 90\% zone occupancy. 
If this threshold is not met, ZenFS delays \texttt{FINISH} by relaxing lifetime matching.
Mixing different lifetimes in one zone increases space amplification, since zones are only reset once all data is invalidated.

\Paragraph{Evaluation Metrics}  
We measure (i)~\textit{device-level write amplification}, (ii)~\textit{space amplification},
(iii)~\textit{wear}, and (iv)~\textit{interference}.

\SubParagraphno{Device-level write amplification} quantifies the additional writes introduced by the device: \( \text{DLWA} = (W_h + W_d) / W_h \), where \( W_h \) is host-issued writes and \( W_d \) is device-issued writes.

\SubParagraphno{Space amplification} quantifies the amount of invalidated data maintained by the system: \( \text{SA} = (W_h + W_i) / W_h \), where \( W_h \) is host-issued writes and \( W_i \) is the amount of invalidated data. We compute \( W_i \) at each timestamp and report SA as the average over the experiment.

\SubParagraphno{Wear} measures how many times each block is erased. \approach{} tracks wear at the storage element level. All blocks of a storage element have the same wear due to \texttt{FINISH} and \texttt{RESET} granularity.

\SubParagraphno{Interference} is defined as the ratio of baseline throughput to the throughput observed during concurrent \texttt{FINISH} operations.
The baseline throughput is measured by running host-issued I/O independently, without any concurrent \texttt{FINISH} operations.


\subsection{\approach{} Wins Across the Board}
We compare \approach{} against a baseline SSD modeled after ZN540 using the ConfZNS++ emulator.
For \approach{}, we only use the configuration where segment is a superblock.
This is because the device has four parallel units and striped writes naturally span all units, resulting in minimal differences across SilentZNS storage elements.
Although we later evaluate fine-grained storage elements using a custom SSD configuration, we first focus on highlighting the benefits of dynamic zone reclamation in baseline SSD.

\begin{figure*}[t]
    \centering
    \includegraphics[width=\textwidth]{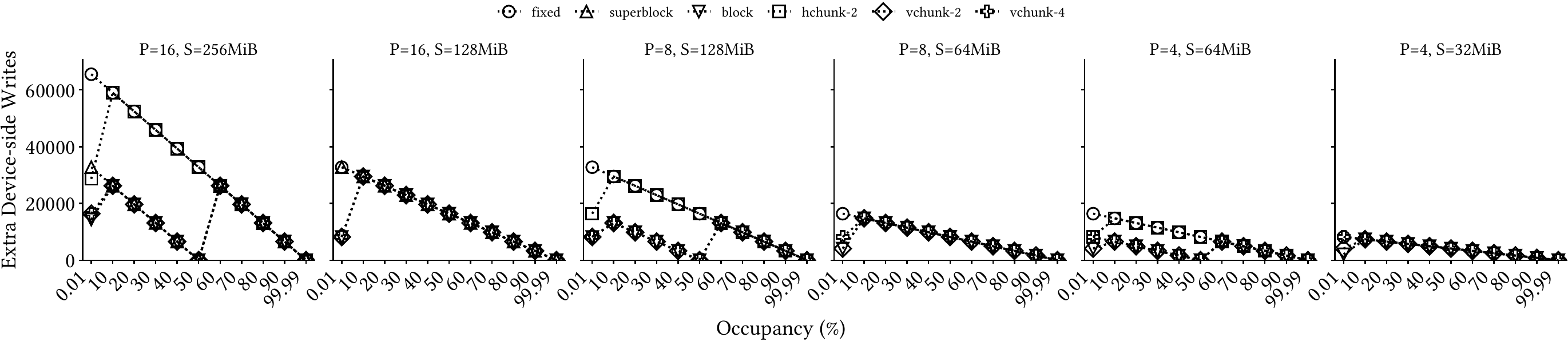}
    \vspace{-6mm}
    \caption{
    Pages finished across different zone geometry configurations (zone parallelism \(P\) and zone size \(S\)).
    Fine-grained allocation schemes (e.g., block and Vchunk) reduce the number of pages that must be finished, especially at lower occupancy.
    }
    \label{fig:pages-finished}
    \vspace{-2mm}
\end{figure*}

\Paragraph{\approach{} Largely Eliminates DLWA}
In our first experiment we use \textit{FINISH benchmark} to evaluate DLWA by filling the zone to occupancy level between 10\% and 90\% before issuing the \texttt{FINISH} command.
Figure~\ref{fig:raw_bench} demonstrates that DLWA is consistently lower for \approach{}, reducing DLWA by up to 86.36\% (10\% zone occupancy with the \textit{superblock} configuration).
This is due to \approach{} writing dummy data only to partially written storage elements while releasing unused ones, therefore, avoiding full-zone dummy writes.
In \approach{} DLWA depends on the position of the write pointer.
If all erase blocks within a segment are fully written, no dummy writes are required.
Consequently, the maximum number of dummy writes is bounded by the size of a segment, since host writes one segment fully before moving to the next.
As a result, at 50\% zone occupancy, approximately half of the segments are fully written with host data, allowing \approach{} to achieve $DLWA = 1$.
In contrast, the baseline approach exhibits DLWA that is strongly correlated with zone occupancy.
It incurs high DLWA at low occupancy due to full-zone dummy writes, whereas \approach{} maintains low DLWA.
At higher occupancy levels, the baseline is competitive as fewer pages remain unwritten.
However, as we show next, delaying \texttt{FINISH} introduces additional costs.

\Paragraph{Reducing SA without Adding DLWA}
In this experiment, we run KVBench-II (4 million total operations with 512B entry size) on RocksDB with ZenFS as the filesystem backend.
We vary the occupancy threshold of ZenFS between 10\% and 90\%.
As Figure~\ref{fig:raw_bench}(b) shows, increasing the threshold and therefore delaying finish increases space amplification, since ZenFS has to relax file lifetime matching requirements.
For instance, at 10\%, the average space amplification is about 1.5, while at 90\% is 2.6.
 
In this experiment we report average SA between basline and \approach{}, as space amplification is independent of the SSD's internal mapping strategy.
Therefore LSM-users can improve SA by lowering the occupancy threshold.
However, this improvement comes at a cost for the baseline SSD.
As shown in Figure~\ref{fig:raw_bench}(b), frequent \texttt{FINISH} commands at low thresholds (which fully write physical zones) result in large amounts of device-side garbage.
In contrast, \approach{} consistently maintains low device-side garbage due to efficient zone management.
At 10\% occupancy threshold, for example, \approach{} has 92\% less DLWA than the baseline.

\Paragraph{Reducing the Unnecessary Wear}
We evaluate erase block wear under \approach{} and baseline using KVBench-II on RocksDB with ZenFS at a occupancy threshold of 10\%.
To capture the cumulative wear, we repeat KVBench-II $8~\times$ due to memory constraints of the SSD emulator.
Figure~\ref{fig:raw_bench}(d) shows that \approach{} reduces unnecessary wear by writing only required storage elements during \texttt{FINISH}, whereas baseline  fills entire zones with dummy writes.
Specifically, our superblock allocation has a total erase count of 15340, compared to 17344 for the baseline.
As Figure~\ref{fig:raw_bench}(d) shows \approach{} also achieves better wear-leveling as erase counts are more evenly distributed across the erase blocks.
This is expected, as baseline ignores wear and allocates the first available physical zone, while \approach{} enforces wear-aware allocation.

\Paragraph{\approach{} Reduces Interference with Host I/O}
We now highlight the benefits of \approach{} for host-issued writes with \textit{Interference Benchmark}.
In this experiment we first fill the zones up to 40\% occupancy.
Then we run the \textit{interference} benchmark with concurrency level ranging from 1 to 7.
As shown in Figure~\ref{fig:raw_bench}(d)
\approach{} consistently maintains lower interference than baseline, with an interference factor of approximately 1.
The baseline experiences interference as high as 1.6 after 4 threads due to contention between host writes and device-issued dummy writes.
Additionally, in our KVBench-II experiment, we observe up to $3.7\times$ faster workload execution with \approach{} (10\% occupancy threshold), as it eliminates unnecessary writes and reduces interference with host I/O.

\subsection{ZNS Design Space Evaluation}
Now we evaluate the ZNS design space and provide guidelines for the host to navigate the variety of ZNS SSD configurations and use them effectively.
We use our custom SSD to study different zone geometries, storage elements and tradeoffs between DLWA, SA, throughput and zone allocation overhead.

\begin{figure*}[t]
    \centering
    \includegraphics[width=\textwidth]{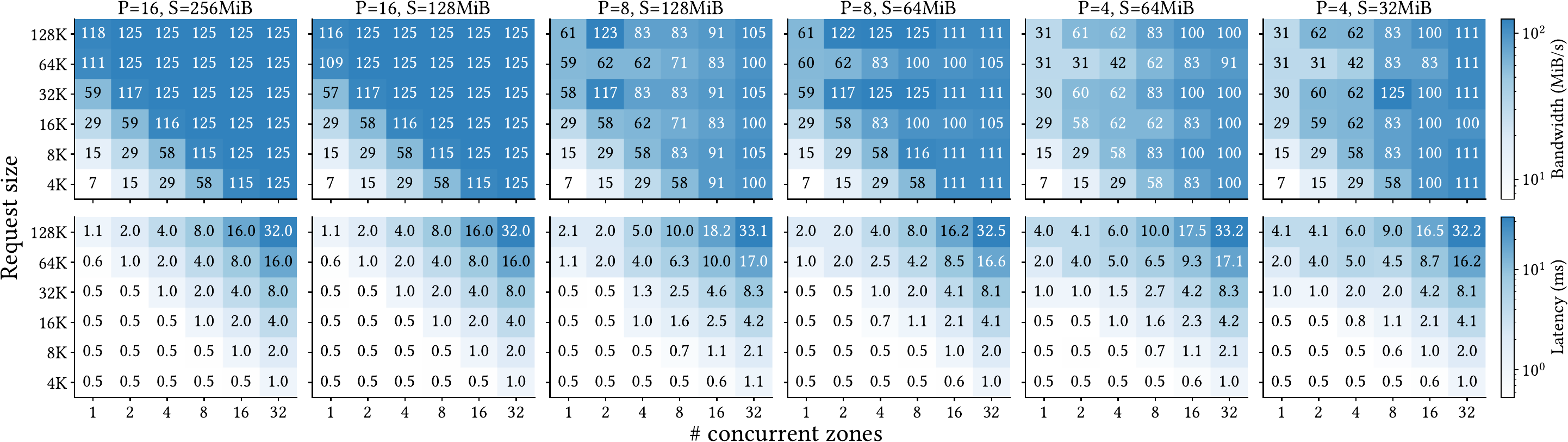}
    \vspace{-6mm}
    \caption{
    Performance across six zone geometries (parallelism \(P\), size \(S\)) under the fixed allocation scheme, showing that larger zones provide higher intra-zone parallelism, while smaller zones require concurrent zones to achieve comparable throughput.
    }
    \vspace{-1mm}
    \label{fig:rw-heatmaps-fixed}
\end{figure*}

\Paragraph{Unnecessary Page Writes Reduce with Small Zones}
In this experiment, we vary the zone size and use \textit{DLWA Benchmark} to fill each zone to occupancy levels between \(0.01\%\) and \(99.99\%\) before issuing the \texttt{FINISH} command. Figure~\ref{fig:pages-finished} shows that for baseline \textit{fixed} zone approach the number of dummy pages written consistently decreases as the zone size becomes smaller. This is because under fixed allocation, \texttt{FINISH} applies to the entire zone regardless of how much data has been written, forcing all unwritten pages to be filled with dummy data. As a result, larger zones incur substantially more unnecessary writes, while smaller zones limit the amount of unwritten space and therefore reduce the need for dummy writes.
This effect is most pronounced at low occupancy. For example, at \(0.01\%\) occupancy, reducing the zone size from 256\,MiB to 128\,MiB reduces the number of dummy writes by approximately \(2\times\), and further reducing zone size to 64\,MiB and 32\,MiB reduces dummy writes \(4\times\) and \(8\times\), respectively. This trend persists across all occupancy levels, with the largest improvements observed at \(0.01\%\) and \(10\%\), where the fraction of unwritten space is highest. As occupancy increases, the benefit of smaller zones diminishes. For instance, at \(90\%\) occupancy, the number of dummy writes is already low across all configurations, and at \(99.99\%\) occupancy it approaches zero.

\vspace{0.5em}
\noindent
\fbox{%
\parbox{0.97\linewidth}{%
\textbf{Guideline G1:}
Prefer smaller zones for workloads that trigger early \texttt{FINISH} (e.g., WAL, OLTP logs or mixed-lifetime data in RocksDB) to reduce DLWA.
}}

\Paragraph{\approach{} Further Reduces Unnecessary Writes}
In addition to reducing the zone size, DLWA can be further reduced by \approach{}. 
The benefits depend on the storage element granularity, zone geometry, and the zone occupancy level.
As shown in Figure~\ref{fig:pages-finished}, \approach{} with \textit{block}, \textit{Vchunk}, and \textit{superblock} provides the most benefit in configurations where a zone spans more than one segment such as \(P=16, S=256\)\,MiB,  \(P=4, S=64\)\,MiB and \(P=8, S=128\)\,MiB.
In these geometry configurations at \(50\%\) occupancy, all \approach{} configurations eliminate dummy writes, as host writes have already completed one segment, and the remaining empty blocks are released for subsequent allocations.
In \approach{}, the upper bound of dummy writes depends on the size of the segment (e.g., a superblock for \(P=16\), and 8 or 4 erase blocks for \(P=8\) and \(P=4\), respectively).
The actual number of dummy pages varies between 0 and these bounds depending on the occupancy level.

When a zone consists of only a single segment (i.e., one superblock or less), \approach{} largely behaves like the baseline fixed allocation, since \texttt{FINISH} must complete all blocks in the segment. 
However, at very low occupancy (e.g., \(0.01\%\)), \textit{block}, \textit{Vchunk-2}, \textit{Vchunk-4}, and \textit{Hchunk-2} still provide gains due to their finer granularity.
For example, in \(P=8, S=128\)\,MiB configuration, \textit{block}, \textit{Vchunk-2}, and \textit{Vchunk-4} reduce dummy writes by approximately \(4\times\) compared to fixed allocation at \(0.01\%\) occupancy.

\vspace{0.5em}
\noindent
\fbox{%
\parbox{0.97\linewidth}{%
\textbf{Guideline G2:}
Use \approach{} to further reduce DLWA, especially when zones span more than one segment. 
At very low occupancy, prefer finer-grained allocation (e.g., \textit{block} or \textit{Vchunk}).
}}

\Paragraph{Managing DLWA vs Throughput Tradeoff}
While reducing the zone size lowers the number of dummy pages written, it introduces a tradeoff with intra-zone scalability.
In this experiment, we use \textit{Write Benchmark} where we vary the request size between 4K and 128K and the number of concurrent zones from 1 to 32.
We identify that if the zone size is reduced while maintaining the same zone parallelism, the system can still achieve peak intra-zone bandwidth, as all parallel units continue to be utilized.
For example, in Figure~\ref{fig:rw-heatmaps-fixed}, both 256 MiB and 128 MiB configurations with zone parallelism of 16 achieve the same peak bandwidth of approximately 110 MiB/s.
In both cases, this peak can be reached with a single thread at 64 KiB request size, without increasing mean request latency.
In this case, the primary disadvantage of reducing the zone size is an increase in the number of zones, which may lead to higher metadata overhead, especially in high-capacity SSDs.
We also identify that if reducing the zone size comes at the cost of reducing zone parallelism, intra-zone parallelism is diminished, which directly impacts bandwidth.
As shown in Figure~\ref{fig:rw-heatmaps-fixed}, configurations with lower parallelism require more concurrent zones to saturate the device.
For example, when zone parallelism is reduced to 8, writing to a single zone achieves only around 60 MiB/s at 64 KiB request size, and 2 concurrent zones are required to increase the bandwidth to 117 MiB/s.
Similarly, when zone parallelism is reduced to 4, a single thread achieves only around 30 MiB/s with 16K request size without increasing the request latency.
In this zone geometry, the system requires up to 16 threads to approach its maximum bandwidth of approximately 100 MiB/s.

\setlength{\belowcaptionskip}{0pt}  
\setlength{\textfloatsep}{2pt}
\begin{table}[t]
\centering
\small
\caption{Interference factor for different zone geometries (parallelism \(P\), size \(S\)) and storage elements.}
\vspace{-4mm}
\label{tab:interference-factor}
\begin{tabular}{lcccccc}
\toprule
Geometry & fixed & sblock & block & Hchunk2 & Vchunk2 & Vchunk4 \\
\midrule
$P4,S32$   & 1.5 & N/A & 1.5 & N/A & 1.5 & 1.5 \\
$P4,S64$   & 1.6 & N/A & 1.1 & 1.6 & 1.1 & 1.1 \\
$P8,S64$   & 1.5 & N/A & 1.6 & N/A & 1.6 & 1.6 \\
$P8,S128$  & 1.6 & N/A & 1.1 & 1.6 & 1.1 & 1.1 \\
$P16,S128$ & 1.6 & 1.6 & 1.6 & N/A & 1.6 & 1.6 \\
$P16,S256$ & 1.6 & 1.1 & 1.1 & 1.6 & 1.1 & 1.1 \\
\bottomrule
\end{tabular}
\end{table}

\vspace{0.5em}
\noindent
\fbox{%
\parbox{0.97\linewidth}{%
\textbf{Guideline G3:}
Use large zones for throughput-critical workloads (e.g., large compactions or bulk ingest); Use smaller zones to reduce DLWA. If reducing zone size lowers parallelism, increase the number of concurrent zones to maintain throughput.
}}

\Paragraph{Fine-grained Storage Elements Reduce Interference}
In this experiment, we run \textit{FINISH} benchmark with a concurrency level of 8.
We identify that interference benefits depend on the storage element and the number of segments in a zone.
As shown in Table~\ref{tab:interference-factor}, SilentZNS with fine-grained allocation (\textit{block} and \textit{Vchunk}) provides the most benefit in configurations where a zone spans multiple segments (e.g., \(P=4,\ S=64\) and \(P=16,\ S=256\)).
In these configurations, interference decreases from 1.6 (under fixed and \textit{Hchunk} allocation) to 1.1, as fine-grained \approach{} configurations (e.g., \textit{Block} and \textit{Vchunk-2}) only write the minimum required storage elements with dummy writes.
However, when a zone consists of a segment (e.g., \(P=4,\ S=32\) and \(P=16,\ S=128\)), all configurations in SilentZNS behave similarly to the baseline \textit{fixed} allocation.
In these cases, \texttt{FINISH} must complete all blocks in the segment, leaving no opportunity to reduce interference, and all schemes exhibit similar interference (e.g., 1.5--1.6).

\begin{table}[t]
\centering
\small
\caption{Median zone allocation latency ($\mu$s) for different zone geometries (parallelism \(P\), size \(S\)) and storage elements.}
\vspace{-4mm}
\label{tab:allocation-latency}
\begin{tabular}{lcccccc}
\toprule
Geometry & fixed & sblock & block & Hchunk2 & Vchunk2 & Vchunk4 \\
\midrule
$P4,S32$   & 0.5 & N/A  & 7991 & N/A  & 6627 & 6283 \\
$P4,S64$   & 0.6 & N/A  & 8247 & 6649 & 6424 & 6026 \\
$P8,S64$   & 0.7 & N/A  & 7995 & N/A  & 6658 & 6014 \\
$P8,S128$  & 0.5 & N/A  & 8475 & 7348 & 7183 & 6108 \\
$P16,S128$ & 0.7 & 1660 & 9030 & N/A  & 6975 & 6358 \\
$P16,S256$ & 0.5 & 1710 & 9068 & 7403 & 7096 & 6936 \\
\bottomrule
\end{tabular}
\end{table}

\begin{table*}[!ht]
\centering
\small
\setlength{\tabcolsep}{4pt}
\caption{Practical guidance for Data Systems based on ZNS design tradeoffs.}
\vspace{-4mm}
\label{tab:guide}
\begin{tabular}{p{3.0cm}p{3.2cm}p{3.6cm}p{6.7cm}}
\toprule
\textbf{DBMS scenario} & \textbf{Recommended setting} & \textbf{Host policy} & \textbf{Takeaway} \\
\midrule

\textbf{(A) WAL / OLTP logs} &
\texttt{block}/\texttt{Vchunk-2},small zones &
Issue early \texttt{FINISH}; isolate WAL zones &
WAL segments are frequently finished at low occupancy; small zones (G1) and fine-grained allocation (G2) minimize DLWA and reduce interference on latency-critical writes. \\

\addlinespace[1pt]
\textbf{(B) LSM flushes / minor compactions} &
\texttt{superblock}/\texttt{Vchunk-4}, medium zones &
Finish at SST boundaries; moderate concurrency &
Flushes may leave partially filled regions; \texttt{Vchunk-4} and \texttt{superblock} balance DLWA and allocation latency (G5) while preserving throughput (G3). \\

\addlinespace[1pt]
\textbf{(C) Large compactions / bulk ingest} &
\texttt{superblock}/\texttt{Vchunk-4}, large zones &
Prioritize throughput; maintain high intra-zone parallelism &
Throughput dominates; preserving parallelism is critical (G3), and coarse granularity is sufficient since DLWA is low at high occupancy. \\

\addlinespace[1pt]
\textbf{(D) Mixed-lifetime ZenFS / RocksDB data} &
block/\texttt{Vchunk-2},small zones &
Issue early \texttt{FINISH} to reduce SA &
\approach{} reduces DLWA and \texttt{FINISH} interference (G2 \& G4), eliminating the need for delaying finish. \\

\addlinespace[1pt]
\textbf{(E) Read-mostly workloads} &
\texttt{superblock}/\texttt{Vchunk-4}, large zones &
Optimize for read parallelism; minimize allocation overhead &
DLWA is less critical; prefer coarse granularity to reduce allocation latency (G5) and maximize throughput. \\

\bottomrule
\end{tabular}
\vspace{-4mm}
\end{table*}

\vspace{0.5em}
\noindent
\fbox{%
\parbox{0.97\linewidth}{%
\textbf{Guideline G4:}
Use \approach{} to further reduce \texttt{FINISH} interference, especially when zones span more than one segment. 
}}

\Paragraph{Managing DLWA vs Zone Allocation Tradeoff}
In this experiment we measure the zone allocation time with different storage elements and zone geometry.
As shown in Table~\ref{tab:allocation-latency}, fixed allocation incurs negligible latency (0.5–0.7\,$\mu$s), while fine-grained strategies introduce higher overhead.
In particular, \textit{block} allocation has the highest cost (up to 9{,}068\,$\mu$s), while \textit{Vchunk-2}, \textit{Vchunk-4} and \textit{Hchunk-2} reduce this overhead to 6{,}000–7{,}000\,$\mu$s.
\textit{Superblock} provides a middle ground, incurring moderate overhead (1{,}600–1{,}700\,$\mu$s).
This latency can be further amortized by pre-allocating and buffering storage elements for future allocations.

\vspace{0.5em}
\noindent
\fbox{%
\parbox{0.97\linewidth}{%
\textbf{Guideline G5:}
Use finer-grained allocation for DLWA-sensitive workloads, and prefer coarser granularity for workloads where DLWA is less critical (e.g., read-mostly)
}}

\section{A Recommendation for ZNS Vendors}

\Paragraph{Navigating Tradeoffs on ZNS Design Space}
Our results highlight fundamental tradeoffs in the ZNS design space across DLWA, interference, throughput, and allocation latency. 
Smaller zone sizes reduce DLWA, especially for workloads that trigger early \texttt{FINISH}, but may require higher inter-zone concurrency to maintain throughput if intra-zone parallelism is reduced. 
\approach{} further reduces unnecessary writes and interference, particularly when zones span multiple segments and at very low occupancy, where finer-grained allocation is most effective. 
However, these benefits come at the cost of higher allocation latency, making coarser storage elements (e.g., \textit{vchunk} or \textit{superblock}) preferable when balancing DLWA and zone allocation overhead.

\Paragraph{The Need for a Flexible Interface}
Overall, there is no one-size-fits-all configuration.
Different workloads may require different zone sizes and storage elements. 
During our experiments, we further observed that at very low occupancy thresholds (e.g., 1\%), both the baseline and \approach{} can fail due to insufficient host-visible address space. 
This is because frequent \texttt{FINISH} operations are triggered at low occupancy.
After a zone is finished, the remaining unwritten LBAs in that zone become unusable, preventing the application from utilizing available empty physical capacity.
While \approach{} can reclaim unused physical space, fixed logical zones limit its ability to fully utilize this capacity.

This motivates the need for a more flexible and configurable zoned interface, where the device exposes multiple configuration options and the host selects an appropriate design at namespace initialization. 
\approach{} enables host applications to explore and evaluate these design choices without modifying the underlying device. 
In this work, we provide metrics and guidelines to help applications reason about the ZNS design space, and to simulate different SSD configurations before deployment.

\vspace{-3mm}
\section{A Practical Guide for Data Systems}
Using the observations and the guidelines from Section~\ref{sec:ZNS-Evaluation}, we now present a practical guide for data systems engineers that can help make the most of ZNS devices, focusing mostly on LSM Stores~\cite{Sarkar2023}.
Table ~\ref{tab:guide} summarizes the practical recommendations detailed below. 



\Paragraph{Use Case 1: Write-Ahead Logging} For database logging, a data system may need to be able to issue an early \texttt{FINISH} command to discard unnecessary parts of a log aggressively. This might lead to high DLWA due to the potential low zone occupancy. To address this, we propose using fine-grained storage elements (e.g., \textit{block} or \textit{Vchunk-2}) that reduce DLWA (G2), and if possible, smaller zone sizes (G1). This design will also reduce zone interference with writes coming from other parts of the data system.

\Paragraph{Use Case 2: LSM flushes and Partial Compactions} LSM memtable flushes \cite{Facebook2021,Kaushik2024} and partial compactions \cite{Sarkar2021c} generate small files of \textit{potentially} similar lifetime and have high concurrency.
In such cases, the DLWA-related benefits of small zones are reduced. 
Here, the recommendation is to have medium-sized zones that allow for both inter-zone and intra-zone parallelism without interference (G3).
As a result, we recommend a \texttt{superblock} or a \texttt{Vchunk-4} setting for storage elements, while SilentZNS would ensure that the overall device wear will be kept minimal -- as shown in Figure~\ref{fig:raw_bench}(c).


\Paragraph{Use Case 3: Major or Full Compactions} When the files to be written are much larger in size, like in major or full compactions \cite{Sarkar2021c}, then the primary goal is to maximize ingestion performance, so the recommendation is to exploit the full parallelism of the device and increase the zone size (G3). Using a superblock is typically sufficient and simplifies zone allocation.

\Paragraph{Use Case 4: Mixing Files with Different Lifetimes} When the workload leads to zones with different lifetime hints, we recommend using smaller zone sizes with fine-grained storage elements, since SilentZNS can offer both low DLWA and SA in this setting.


\Paragraph{Use Case 5: Read-Heavy Workloads} 
For read-heavy workloads, DLWA is less critical, and we focus on maximizing read throughput and minimizing allocation overhead. 
To achieve this, we recommend large zone sizes with coarse allocation granularity (G5).

Overall, our analysis informs data systems and highlights the need for \textit{bounded} flexibility from device vendors, so that data-intensive systems can make the most of the underlying ZNS SSDs. 


\section{Conclusion}

ZNS SSDs provide a new append-only interface that eliminates device-side garbage collection and provides stable throughput. 
However, existing designs suffer from high DLWA, which causes increased wear, interference with host I/O, and a need to balance a complex SA vs DLWA tradeoff.
To address this, we design and evaluate a spectrum of flexible zone allocation strategies that allocate only the necessary storage elements on demand and apply management commands like \texttt{RESET} and \texttt{FINISH} at the granularity of these elements.  We integrate \approach{} into the ConfZNS++ SSD emulator and show that \approach{} reduces DLWA by up to $92\%$, while maintaining SA at 1.42 and achieving 3.7~$\times$ faster workload latency. We conclude by sharing recommendations to ZNS SSD vendors and data systems engineers who use such devices.
\balance

\bibliographystyle{ACM-Reference-Format}
\tiny
\bibliography{library}
\balance

\end{document}